# Nonexponential Relaxation of Photoinduced Conductance in Organic Field Effect Transistor

Soumya Dutta and K. S. Narayan[*]

Chemistry and Physics of Materials Unit,

Jawaharlal Nehru Centre for Advanced Scientific Research,

Jakkur Campus, Bangalore 560 064, India

(Dated: February 3, 2003)

We report detailed studies of the slow relaxation of the photoinduced excess charge carriers in organic metal-insulator-semiconductor field effect transistors consisting of poly(3-hexylthiophene) as the active layer. The relaxation process cannot be physically explained by processes, which lead to a simple or a stretched-exponential decay behavior. Models based on serial relaxation dynamics due to a hierarchy of systems with increasing spatial separation of the photo-generated negative and positive charges are used to explain the results. In order to explain the observed trend, the model is further modified by introducing a gate voltage dependent coulombic distribution manifested by the trapped negative charge carriers.

PACS numbers: 85.60.Dw, 71.20.Rv, 72.20.Jv, 72.80.Le

Field effect transistors based on organic semiconductors and conjugated polymers have spurred tremendous interests in the development of organic/plastic electronics[1, 2] and opto-electronics[3, 4] due to the lower cost, facile processability and chemical tenability. Apart from the technological implications, the metal-insulator-semiconductor field effect transistor (MISFET) geometry can also serve as a model structure to study the charge transport mechanisms and transport parameters such as field effect mobility, and semiconductor-insulator interfacial density of trap states. The selective distribution of the carrier density in the FET geometry with the large concentration of the positive charges in the channel region compared to the bulk can result in interesting situation especially upon photoexcitation. Large changes in the dark drain-source current $I_d$ or $I$ upon photoexcitation, at low-light levels was recently demonstrated in FETs based on regio-regular poly(3-octylthiophene).[5]

It was observed that in an organic FET, which is typically p-channel and activated under enhancement mode, the incident light intensity can act as an added control parameter.[5] The electric field distribution in the device promotes the separation of the photoinduced charge carrier (PCC). The more mobile positive charges drift towards the channel whereas the negative charges find their minimum potential in the bulk and get localized in the trap sites. Upon, switching off the photoexcitation, the photoinduced current $I^{light}(t)$ relaxes quite slowly with an apparent persistent behavior and the sample recovers to the dark conditions over a period of several hours. This persisting effect can lead to back-gating problems in polymer FETs. The decay mechanism in the present system depends on several aspects like defects in insulator-semiconductor interface, bulk trap densities, diffusion rates of PCC and $V_g$. We examine this decay closely as a function of gate voltage ($V_g$), which largely involves the neutralization of the excess negative charges in the bulk in the background of the channel current.

Slow relaxation of PCC is known to occur in poly(phenylenevinylene) (PPV) and poly(alkylthiophene) systems.[6,7,8,9] The non-equilibrium conductance following photoexcitation in these systems was interpreted in terms of dispersive diffusion mechanism restricting recombination. The decay mechanisms are typically fitted to a stretched exponential behavior (Kohlrausch's law); The conductivity $\sigma(t) = \sigma_0 \exp[-(t/\tau)^\gamma]$, where $\tau$ is the relaxation lifetime and the exponent $\gamma$ is indicative of the degree of disorder and associated with a distribution of activation energies. Two different statistical approaches have been used to describe the decay processes. One proposed on the basis of a statistical distribution of lifetimes of many correlated degrees of freedom (parallel relaxation) and the other based on the serial relaxation to equilibrium essentially through a hierarchy of dynamical constraints proposed by Anderson *et al.*[10] and further modified by Queisser *et al.*[11] We highlight the validity of the latter model and its specific applicability to study $I^{light}(t)$ of the transistor structure in this report.

In conjugated polymers having nondegenerate ground state, charge injection onto the backbone causes a local structural distortion and the formation of self-localized polarons or bipolarons with associated electronic states in energy gap $E_g$.[9] It is also well known that for such system the transport in the bulk mainly governed by bipolaron, which is stable excitation and near the surface bipolaron dissociates into polarons leading to recombination.[6] It has been shown using charge modulation spectroscopy and electroabsorption that the electric field induced charge carriers in the regio-regular

poly(3-hexylthiophene) (RRP3HT) channel and the majority carriers after the photo-generation process are largely polaronic in nature.[12]

After termination of illumination, the decay of $I^{light}(t)$ in semiconducting polymers, such as PPV and its derivatives[6, 7] is typically subjected to a fit with Kohlrausch's law, and described by a relaxation process involving a series of sequential, correlated activation steps.[10] It is observed that the decay rate is characterized by large time constants and $0 < |\gamma| < 1$ in systems especially with long-lived decay ($\gamma \approx 0.5$ for PPV [6]).

The stretched-exponential decay law (SEDL) can also be derived introducing a power-law dependence of decay rate based on continuous-time random walk.[13] The use of the model can lead to interpretation which demands that the physical nature of recombination changes with time, spanning over large timescales. A forcible fit to this model would also then lead to assuming a wide spread in trap energies and cross-sections. A model based on different spatial separation is more appropriate in the present case, which relies on identical physical mechanism for recombination for the rapid and slow segments of decay.[14] The slow process of the decay forms a large component of the entire decay in the FET geometry and makes it necessary to resort to such an approach. The model invoked to explain the present results, which exhibit extremely slow decay process, has been adapted from the model[11] to describe the persistent photoconductivity results in *n*-type gallium arsenide (*n*-GaAs) grown on insulating chromium doped GaAs (Cr: GaAs) with hole traps, where the decay is as follows:

$$\sigma_{\text{off}}(t) = \sigma_{\text{ini}} - \frac{1}{2x_p} e\mu_h aZ \ln(t/\tau_0), \quad \text{for } t \gg \tau_0 \tag{1}$$

where $e$ is the charge of the majority carriers having mobility $\mu_h$ within the conducting channel of thickness $x_p$, $Z$ is the volume density of traps, $\tau_0$ is carrier life time of uniform bulk material and $a$ is the Bohr radius of the positive charge carriers. The equation is derived on the basis of the recombination process of PCC at a rate expressed as[14]

$$\frac{dn(x,t)}{dt} = -\left[n(x,t)/\tau_0\right]\exp(-2x/a) \tag{2}$$

where $n(x,t)$ is trapped negative charge density persisting after time interval $t$ from termination of the disequilibrating illumination. The $x$-axis extends toward the bulk and $x = 0$ corresponds to a discrete channel-bulk boundary. By solving Eq. (2) for $n(x,t)$ in terms of $t$ and assuming $t \gg \tau_0$, which is valid for many practical purposes, the excess conductivity due to PCC can be expressed as

$$\sigma_{\text{off}}(t) x_p = \mu_h e \int_0^\infty n(x,t)\,dx \approx \mu_h e \int_0^\infty n(x,0)\,dx \tag{3}$$

where $n(x,0)$ corresponds to the density of trapped negative charges at $t = 0$. In disordered semiconductors, the less mobile PCC get localized and trapped in the bulk maintaining a distribution after being photoexcited.[14] In present system a rectangular distribution of trapped negative charge carriers is considered to be extended up to a certain length $x = l$ at time $t = 0$, i.e. $n(x,0) = Z$ in the interval $[0, l]$ and zero elsewhere. At time $t$, all carriers having life time $t+\tau_0$ can be assumed to have recombined. The effect of recombination can be realized clearly by considering a sharp-front separating the localized negative charges from the mobile positive charges is moving along *x*-axis. The position of the front relates the time through a spatial distribution $\tau = \tau_0 (2x/a)$, which can be obtained from the denominator of Eq. (2). The lifetime $t + \tau_0$ corresponds to the position of the front $x(t) = (1/2) a \ln [1 + (t/\tau_0)]$, which can be treated as the lower limit of Eq. (3) to find the conductivity. Integration of Eq. (3) by substituting the rectangular distribution finally yields Eq. (1) for $t \gg \tau_0$. The concept of spatial distribution of the carriers, used in this model, is justifiable specifically for FET geometry under photoexcitation. The $V_g$ dependency of $I^{light}(t)$ decay can be taken to be a measure of the parameter governing the spatial distribution. The associated physical arguments arise from a possible presence of an in-built field, which modifies the transistor characteristics, and the charge carrier generation.[15, 16] Upon switching off the light source the field redistribution and the recombination causes the current decay. The rate of recombination decreases progressively accompanied by the relaxation of the in-built field. In the longer time scale the decay process is governed by the recombination of charge carriers separated by larger distances. The thermal relaxation related mechanisms are neglected in the present model, hence low-light intensity and low temperature were employed in the experiments.

The top-contact geometry was used to fabricate the MISFET devices as shown schematically in right inset of Fig. 1. Polyvinyl alcohol (PVA) ($\varepsilon \approx 8$) was spin coated on top of aluminum coated glass substrates as the dielectric layer with thickness $\approx 0.5$ μm.[5] After a thermal treatment process hexamethyldisilazane was coated on the dielectric material and cured for an hour to enrich the surface quality. A layer of RRP3HT (thickness $\approx 150$ nm, $E_g \approx 1.8$ eV) was spin coated onto the dielectric followed by a thermal treatment under vacuum at 60° C for 2 hours. The drain and source electrodes were then fabricated by thermally evaporating gold through a shadow mask to maintain their separation of 40 μm and width of 3 mm.

The transistor characteristics of the MISFET device consisting of PVA as insulating layer and RRP3HT as semiconducting layer, is shown in Fig. 1. In order to analyze the characteristics, the traditional FET model is typically modified to include the bulk-linear effects.[17] Hence the total current in this device is:

$$I_d = I_{channel} + I_{bulk} = \frac{\mu_h C W}{2L}\left[(V_g - V_0)V_{ds} - \frac{V_{ds}^2}{2}\right] + \left(\frac{Ne\mu_h WT}{L}\right)V_{ds} \quad (4)$$

where $C$ is the capacitance per unit area of the dielectric medium (i.e. $C = \varepsilon/d$, $d$ is the thickness of the insulator having dielectric constant $\varepsilon$), $W$, $L$, and $T$ are the channel width, channel length and the thickness of the active material respectively and $N$ is the charge density in the bulk. The experimental data are well-fitted (standard error < 0.3%) to Eq. (4) (dashed lines) providing $\mu_h \approx 2.6 \times 10^{-3}$ cm$^2$/V s at $V_g = -10$ V and $N \approx 10^{15}$/ cm$^3$. The transfer characteristic of that device at $V_{ds} = -10$ V is depicted in left inset of Fig. 1.

The device characteristics are altered dramatically upon photoexcitation. Large changes in $I$ similar to the FETs based on RRP3OT [5] are observed upon photoexcitation. The response of $I^{light}(t)$, measured across drain-source with respect to the photoexcitation are observed to be dependent on $V_g$ as shown in Fig. 2(a). Fig. 2(a) depicts the sharp increase in $I^{light}(t)$ upon exposing to light having energy ≈ 2.3 eV (> $E_g$) and intensity ≈1 mW/cm$^2$ for a device biased at $V_{ds} = -40$ V at 100 K. The initial stages of the decay in the sub millisecond range can be evaluated from the ac response of $I(t)$ to short periodic light pulses as shown in Fig. 2(a) inset. Two distinct features were observed from these experiments: (i) The ac peak-peak response of $I(t)$ is at least three orders less than the total change of dc current obtained from long duration light-exposure. (ii) The ac response, which is a signature of fast response (< 20 msec) does not show any discernible dependence on $V_g$. The $V_g$ dependence of the current decay appears at longer time scales when the electrons and holes are well-separated from each other. The difference in $V_g$ dependence can then be attributed to a more effective screening at higher density of PCC, which prevails at the early stages.

The build up of photoinduced excess current $I_{on}(t)$ for different $V_g$ is shown in Fig. 2(b). The value of $I_{on}(t)$ exhibits logarithmic dependence of time as fitted (dashed lines) in Fig. 2(b). This behaviour was reproduced for devices prepared using polymers from different batches indicating the general validity of this build-up over a measurement time range t < 20000 sec as shown in Fig. 2(b) inset. According to the model given by Queisser *et al* [14] the logarithmic dependence of $I_{on}(t)$ on cumulative photon dose implies a spatial separation of PCC at the interface. The interface potential between the *p*-layer and the bulk causes the separation of PCC with the negative charge diffusion to the substrate and get localized in trap states. The explanation of the accruing conductance was attributed to the available trap sites. Another particular feature observed in the present system is that $I_{on}(t)$ rises faster with increasing negative $V_g$. This can be primarily attributed to the higher drift/diffusion rates of the photo-generated positive charge from the bulk to the channel region where the potential of the positive charge is significantly lowered. The negative charge carriers would also tend to diffuse into the bulk and the spatial-access can increase with increasing $V_g$ in the enhancement mode.

The study of the decay rates upon switching off the PCC generation source is more informative and revealing, especially in terms of the relaxation processes prevailing in the system. The decay clearly, as shown in Fig. 3 reveals the presence of a persistent current, i.e. the photoinduced effect persists for several hours prior to reaching its dark equilibrium value as shown in inset (b) of Fig. 3. The indicators for the dark equilibrium conditions are the saturation current and mobility values obtained from the transistor characteristics.

The current decay $I_{off}(t)$ can be fitted forcefully to the Kohlrausch's law yielding physically, unrealistic value of fitting parameter such as $\gamma = 0.0538$. The present model, however, predicts that $I_{off}(t)$ depends on the density of the excess charge generated in the device during exposure. The fitting to this model using Eq. (1) is satisfactory for the initial duration ($t < 200$ s), at longer duration a small but discernable deviation to the model appears. The discrepancy arises due to our initial assumption based on the rectangular distribution of trapped negative charges, which does not portray a realistic distribution of the spatial of the negative charge carriers in the bulk. A more appropriate and physically explainable distribution can be of the form $n(x,0) = Q/x$ within the region $0 < x \leq l$ and zero elsewhere, where $Q$ is the $V_g$ dependent parameter expressing the density of the negative charges. On substitution of this distribution along with the same limit as described earlier and assuming $t \gg \tau_0$, Eq. (3) leads to

$$I_{off}(t) = I_{ini} - I_{dec} \ln[\ln(t/\tau_0)] \quad (5)$$

where $I_{dec}$ is linearly proportional to $\mu_h$, $a$ and $Q$, governing the decay rate. The experimental points are fitted with Eq. (5) over the entire duration with a higher resolution (standard error < 0.3%), as shown in Fig. 3 (dashed lines). The decay also exhibits a dependence on $V_g$ supporting the distribution introduced in the model.

The $V_g$ dependence of the fitting parameters are depicted in inset (a) of Fig. 3. The parameter $I_{ini}$ represents the initial value of $I_{off}(t)$ prior to the decay, or the saturated photoinduced current. This value is higher for $V_g < 0$ and decreases linearly as the FET is operated in the depletion mode, as one would expect intuitively. The second parameter $I_{dec}$ follows a similar trend with respect to $V_g$. This dependence clearly indicates that the excess charge density and its decay rate depend on $V_g$. The $V_g$ dependence of the decay rate becomes more evident upon examining the derivative of Eq. (5) and further validates the model adapted.

Further insight into the physical aspects of the decay-model in case of the present system were obtained by carrying out a series of experiments, revealing the minimal role of the drain/source contacts and the region around it. For example the polarity of the contact was reversed periodically to sweep out the light-induced trapped charges around the region. This procedure resulted only in marginal changes in the magnitude of the residual current without any significant changes in the temporal aspect of the decay. Similar experiments were also done with light incident selectively on the source/drain electrode region. The persistent behavior and the general profile of the decay existed even after continual flipping of $V_{ds}$. These observations are indicative of relaxation across the bulk-interface region.

In conclusion, an illustrative experimental investigation with a suitable theoretical approach is accomplished to interpret the slow relaxation of the photoinduced current in organic FET. We observe a $V_g$ dependence of photocurrent relaxation in the intermediate stages of the decay in this three terminal device. The model based on the basis of dispersive diffusion of charge carriers after being generated was adapted for the present system. The model was modified to include a gradually decreasing function for the spatial distribution of the photo-generated trapped negative carriers. The model yields excellent fits to the experimental results with reasonable description of the physical attributes in the decay process, especially in the intermediate and long time regimes.

*E-mail: narayan@jncasr.ac.in

[1] C. D. Dimitrakopoulos and D. J. Mascaro, IBM J. Res. Dev. **45**, 11 (2001).

[2] C. J. Drury, C. M. J. Mutsaers, C. M. Hart, M. Matters, and D. M. de Leeuw, Appl. Phys. Lett. **73**, 108 (1998).

[3] A. R. Brown, A. Pomp, M. Hart, and D. M. de Leeuw, Science **270**, 972 (1995).

[4] A. Dodabalapur, Z. Bao, A. Makhija, J. G. Laquindanum, V. R. Raju, Y. Feng, H. E. Katz, and J. Rogers, Appl. Phys. Lett. **73**, 142 (1998).

[5] K. S. Narayan and N. Kumar, Appl. Phys. Lett. **79**, 1891 (2001).

[6] C. H. Lee, G. Yu and A. J. Heeger, Phys. Rev. B **47**, 15543 (1993).

[7] B. Dulieu, J. Wéry, S. Lefrant, and J. Bullot, Phys. Rev. B **57**, 9118 (1998).

[8] M. Westerlin, R. Österbacka, and H. Stubb, Phys. Rev. B **66**, 165220 (2002).

[9] Y. H. Kim, D. Spiegel, S. Hotta, and A. J. Heeger, Phys. Rev. B **38**, 5490 (1988).

[10] R. G. Palmer, D. L. Stein, E. Abrahams, and P. W. Anderson, Phys. Rev. Lett. **53**, 958 (1984).

[11] H. J. Queisser, Phys. Rev. Lett. **54**, 234 (1985).

[12] P. J. Brown, H. Sirringhaus, M. Harrison, M. Shkunov, and R. H. Friend, Phys. Rev B **63**, 125204 (2001).

[13] M. F. Shlesinger and E. W. Montroll, Proc. Nat. Acad. Sci. U.S.A. **81**, 1280 (1984).

[14] H. J. Queisser and D. E. Theodorou, Phys. Rev. B **33**, 4027 (1986).

[15] E. J. Meijera, C. Tanase, P. W. M. Blom, E. van Veenendaal, B. H. Huisman, D. M. de Leeuw, and T. M. Klapwijk, Appl. Phys. Lett. **80**, 3838 (2002).

[16] U. Albrecht and H. Bässler, Chem. Phys. Lett. **235**, 389 (1995).12

[17] F. Garnier, G. Horowitz, D. Fichou, and X. Peng, in *Science and Application of Conducting Polymers*, edited by W. R. Salaneck, D. T. Clark, E. J. Samuelsen (Adam Hilger, New York, 1991), p.73.

**Figure Caption:**

**Fig. 1:** Transistor characteristic of the device with $L$ = 38 μm; $W$ = 2.3 mm; $T$ = 150 nm and $d$ = 0.5 μm. The dashed lines indicate fits to $I_d(V_{ds})$. Inset: (Right) The schematic diagram of a MISFET in a staggered top-contact geometry. (Left) Transfer characteristics at $V_{ds}$ = - 10 V.

**Fig. 2:** (a) The photoinduced current $I^{light}(t)$ in a FET for different $V_g$. The device was exposed with a laser (≈ 2.3 eV) under $V_{ds}$ = -40 V at 100 K. Inset: short pulse (20 ms) response $I(t)$ in ac mode at $V_{ds}$ = 60 V for different $V_g$; -60 V (o), 0 V (Δ) and 60 V (∇). (b) The build up of photoinduced current $I_{on}(t)$ for different $V_g$, along with fits to logarithmic time dependence (- - -). Inset: $I_{on}(t)$ for a device from a different batch under $V_{ds}$ = -80 V, $V_g$ = -20 V at 150 K over a longer time regime (every 500$^{th}$ data point is displayed).

**Fig. 3:** Log-log plot of photocurrent relaxation $I_{off}(t)$, along with fits to Eq. (5) (- - -) at different $V_g$; -32 V (●), -24 V (⊕), -16 V (+), -8 V (Δ), 0 V (∇), 16 V (×), every 70$^{th}$ data point is plotted. Inset (a) Decay fitting parameters; $I_{ini}(V_g)$ and $I_{dec}(V_g)$ as a function of $V_g$, and Inset (b) is the semilog plot of $I_{off}(t)$ under $V_{ds}$ = -80 V, $V_g$ = -20 V at 150 K for device prepared from a different batch over a longer time regime along with SEDL fit (—) and double-log fit (- - -).

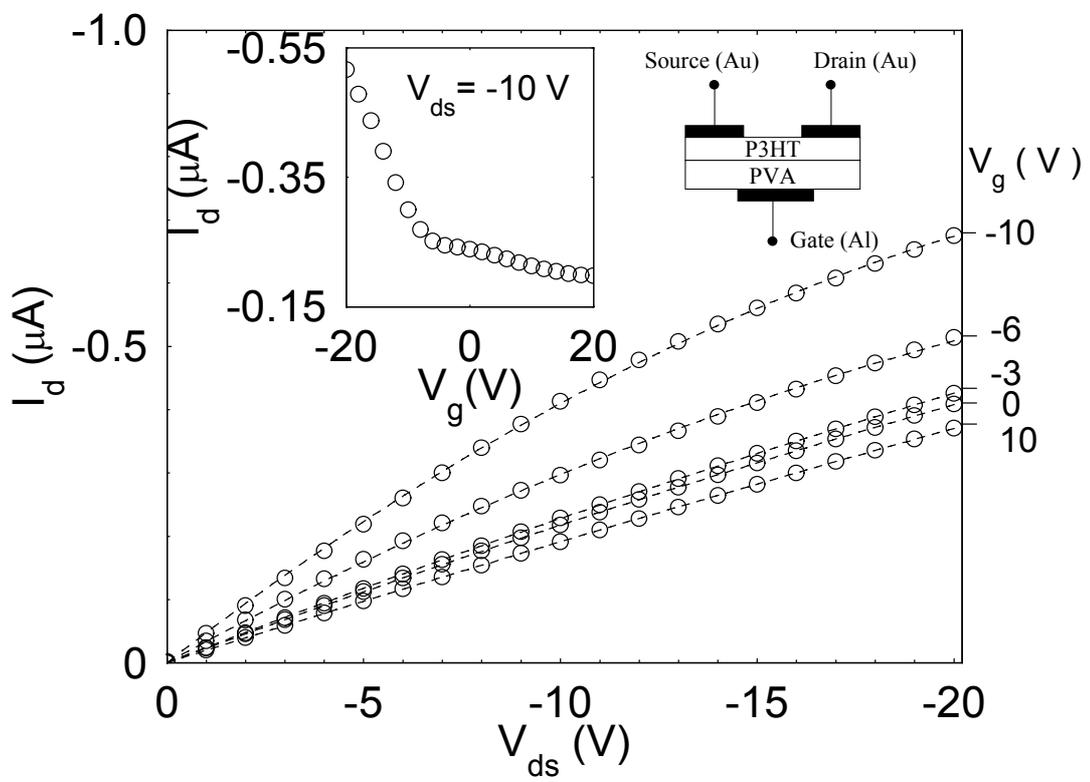

Fig. 1

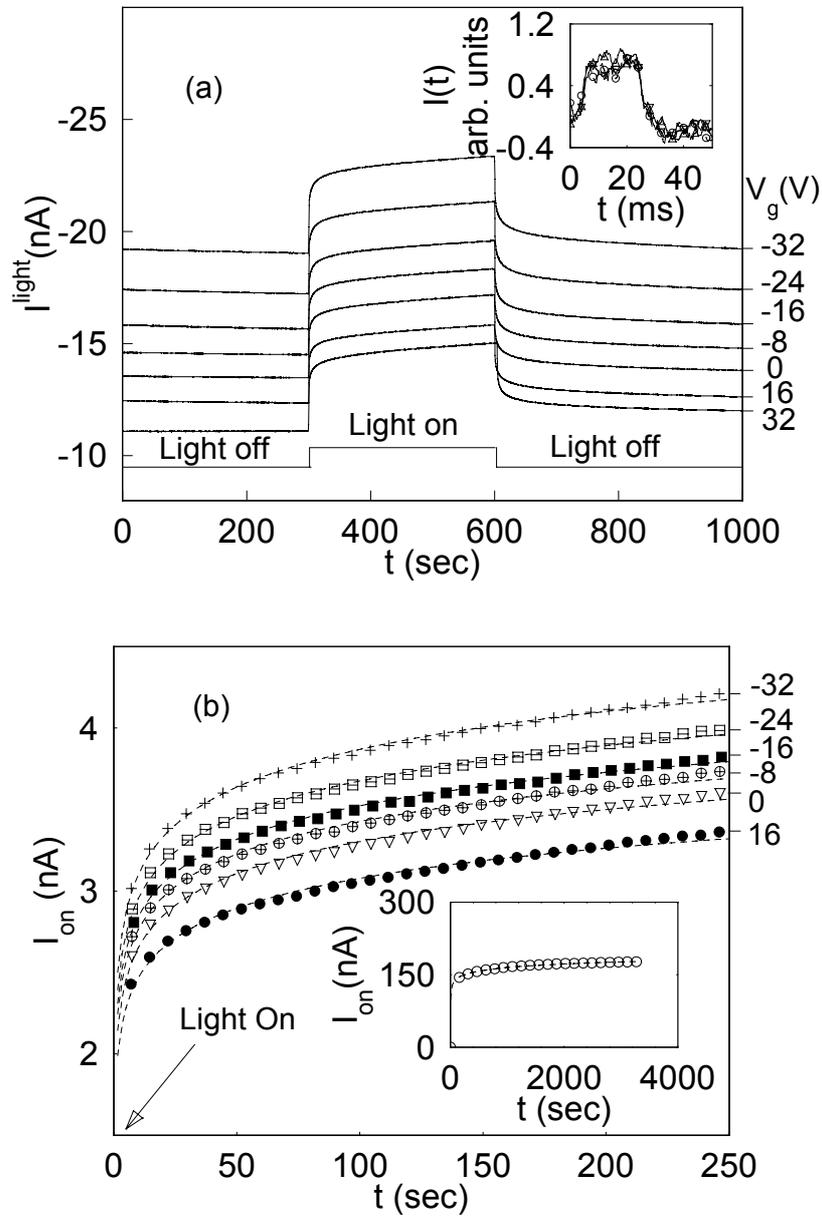

Fig. 2

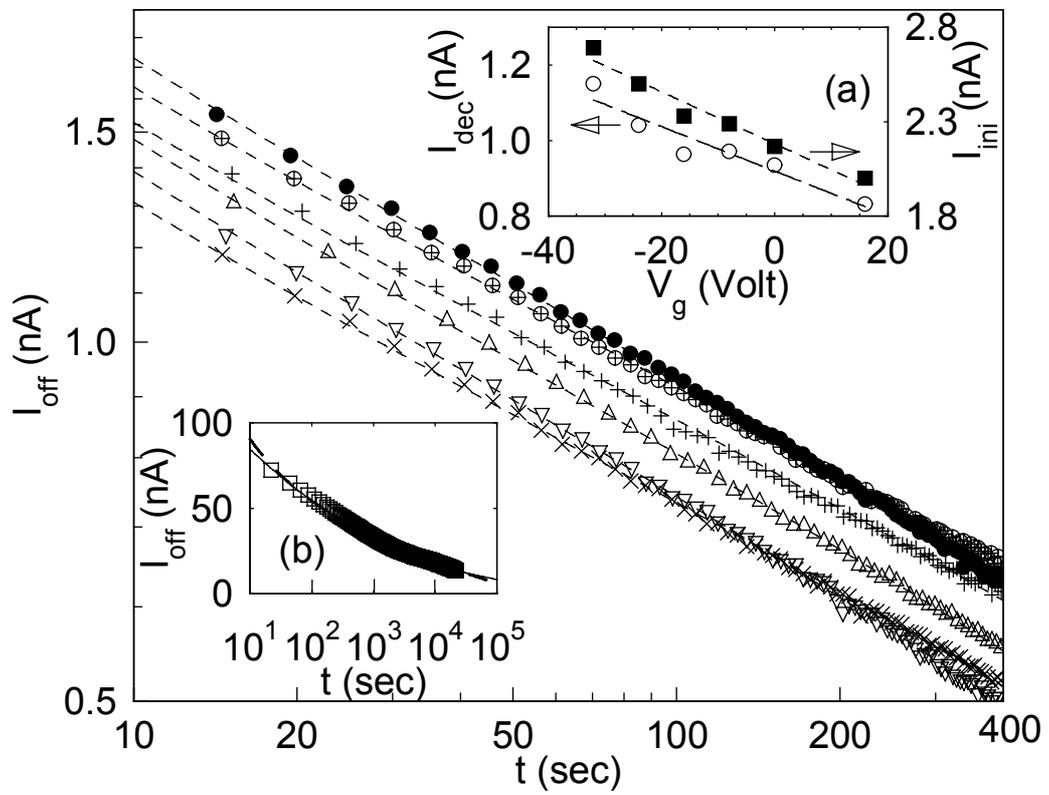

Fig. 3